\documentstyle[stwol,epsfig]{article}

\def\NPB{{\em Nucl. Phys.} B}
\def\PLB{{\em Phys. Lett.}  B}

\def\PRD{{\em Phys. Rev.} D}
\def\ZPC{{\em Z. Phys.} C}

\def\be{\begin{equation}}
\def\ee{\end{equation}}
\def\bea{\begin{eqnarray}}
\def\eea{\end{eqnarray}}

\def\sqee{\sqrt{s}_{\rm ee}}

\def\gg{\gamma\gamma}
\def\Wvis{W_{\rm vis}}

\def\qmax{Q^2_{\rm max}}

\def\ee{\mbox{e}^+\mbox{e}^-}

\def\WECAL{W_{\rm ECAL}}

\newcommand{\sleq} {\raisebox{-.6ex}{${\textstyle\stackrel{<}{\sim}}$}}

\def\ETJET{E^{\rm jet}_T}

\def\ETi{E_{T_i}}

\def\qqbar{q\overline{q}}
\def\xg{x_{\gamma}}
\def\xgp{x_{\gamma}^+}
\def\xgm{x_{\gamma}^-}
\def\xgpm{x_{\gamma}^{\pm}}
\def\etajet{\eta^{\rm jet}}
\def\Zzero{\ifmmode {{\mathrm Z}^0} \else {${\mathrm Z}^0$} \fi}
\def\ppbar{\overline{\mbox p}\mbox{p}}

\def\phijet{\phi^{\rm jet}}

\begin{document}

\renewcommand{\thefootnote}{\fnsymbol{footnote}}



\title{JET PRODUCTION IN PHOTON-PHOTON INTERACTIONS \footnotemark[1]}
\thispagestyle{myheadings}
\markright{FREIBURG-EHEP-96-01}

\author{\vskip -2.5 mm STEFAN S\"OLDNER-REMBOLD}

\address{Albert-Ludwigs-Universit\"at Freiburg,
Fakult\"at f\"ur Physik, D-79104 Freiburg im Breisgau, Germany}

\author{\vskip -1mm On behalf of the OPAL collaboration}


\twocolumn[\maketitle\abstracts{
The inclusive one- and two-jet cross-sections are
measured in collisions of quasi-real photons at 
e$^+$e$^-$ centre-of-mass energies $\sqee=130$ and $136$ GeV using the OPAL
detector at LEP. Jets are reconstructed with a cone
jet finding algorithm. The jet cross-sections are compared to
next-to-leading order (NLO) perturbative QCD calculations.
Transverse energy flows in jets are studied separately for direct and resolved
two-photon events. 
}]

\section{Introduction}
\footnotetext[1]{To be published in the proceedings of ICHEP'96, Warsaw}
We present measurements of jet production in $\gg$ interactions 
at $\sqee$ of 130 and 136 GeV.
The jet cross-sections are compared to
predictions of perturbative QCD which use different
parametrisations of the photon structure function.

In the Quark Parton Model (QPM) jets are produced 
by the interaction of bare photons, $\gg\rightarrow \qqbar$.
This is called the direct process. 
The largest part of the total cross-section, however,
is modelled by interactions where the photon fluctuates
into a hadronic state. The processes are called single-resolved if one photon
couples directly to a parton in the other photon
and double-resolved if partons from both photons interact. 

\section{Data analysis and \\ Monte Carlo simulation}
\label{sec-data}
Two-photon events are selected by requiring that
the sum of all energy deposits in the calorimeters
is less than 50 GeV.
The hadronic invariant mass, $\WECAL$, measured
in the electromagnetic calorimeter has to be greater than 3 GeV;
the missing transverse energy of the event
has to be less than 5 GeV; at least 5 charged tracks must have 
been found and no track may have a
momentum greater than 15 GeV/$c$.

To remove events with scattered electrons in the forward detectors,
the energy measured in these detectors has to be less than
40 GeV (20 GeV) depending on the angular range,
corresponding to an effective maximum
photon virtuality, $\qmax$, of 0.8~GeV$^2$ (``anti-tagging condition'').

For an integrated luminosity of
4.9 pb$^{-1}$, 7808 events remain after all cuts. 
The average visible hadronic invariant mass $\Wvis$ is about 18 GeV.
The total background from other processes is about 1~\%.

\begin{figure}[htbp]
   \begin{center}
      \mbox{
\hskip -0.1cm
          \epsfxsize=7.58cm
           \epsffile{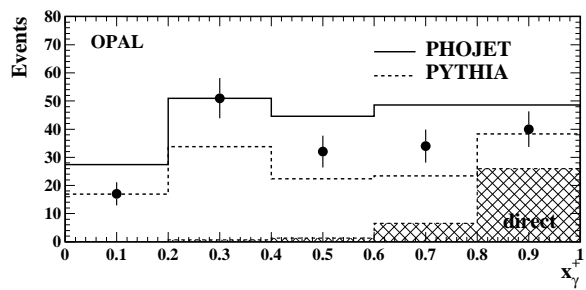}
           }
   \end{center}
\caption{The number of two-jet events
as a function of $\xgp$ compared
to PHOJET and PYTHIA. 
Statistical errors only are shown.
The hatched histogram is the
direct contribution to the PYTHIA events.
}
\label{fig-xg}
\end{figure}

The leading order (LO) QCD generators PYTHIA 5.721~\cite{bib-pythia} and 
PHOJET 1.05~\cite{bib-phojet} are used for the $\gg$ event simulation.
The fragmentation is handled by JETSET 7.408~\cite{bib-pythia}.

\section{Jet reconstruction and backgrounds}
\label{sec-cone}
In the cone jet finding algorithm a jet is defined as a set of 
particles, i.~e.~partons or hadrons generated in a Monte Carlo program, 
or tracks and calorimeter clusters,
whose momenta lie within a cone of size
$R$, such that the axis of the cone coincides
with the momentum sum of the particles contained. 
The cone size $R$, which is chosen to be $R=1$, is defined as
$R=\sqrt{(\Delta\eta)^2+(\Delta\phi)^2},$
with $\eta=-\ln\tan(\theta/2)$ being the pseudorapidity and $\phi$ the
azimuthal angle in the laboratory frame. 
$\Delta\eta$ and $\Delta\phi$ are the differences
between the cone axis and the particle direction.
The total transverse energy $\ETJET$ of the jet 
(defined with respect to the beam axis)
is the scalar sum of the transverse energies $\ETi$ of its components.
The pseudorapidity $\etajet$ and the
azimuthal angle $\phijet$ are defined as the sum over
the $\eta_i$ and $\phi_i$
of the jet components weighted by their transverse energies $\ETi$:
$$\etajet=\frac{\sum_{i} \ETi \eta_{i}}{\sum_{i} \ETi}
\;\;\;\;\mbox{and}\;\;\;
\phijet=\frac{\sum_{i} \ETi \phi_{i}}{\sum_{i} \ETi}.$$
All results are given for jets with $\ETJET >3$~GeV and
$|\etajet|<1$.
\markright{ }
\section{Energy flow in jet events}
\label{sec-flow}
In $\gamma$p scattering, direct and resolved two-jet events,
which correspond to the single-
and double-resolved events in $\gg$ scattering if the
proton is substituted in the place of a VMD-like photon,
have been identified by measuring the fraction $\xg$ of
the photon energy participating in the hard scattering~\cite{bib-xghera}. 
In $\gg$ scattering two photons interact,
therefore a pair of variables is defined~\cite{bib-LEP2}
\begin{equation}
\xgpm=\frac{\sum_{\rm jets}(E\pm p_z)} {\sum_{\rm hadrons}
(E\pm p_z)}, 
\label{eq-xgpm}
\end{equation}
where $p_z$ is the momentum component along the $z$ axis of the
detector and $E$ is the energy.
Ideally, the total energy of the event is contained in the 
two jets for direct events without remnant jets, i.~e.~$\xgpm=1$,
whereas for resolved events at least one of the $\xgpm$ values
is expected to be much smaller than one.
The variables $\xgpm$ are measured using all tracks and 
calorimeter clusters that were used 
in the jet finding algorithm. In addition,
the energies in the forward detectors, which are not used for
jet finding and for the energy flow distributions, 
are added in the denominator of Eq.~\ref{eq-xgpm}.
This improves the separation of direct and resolved events.

\begin{figure}[htbp]
   \begin{center}
\hskip -0.15cm
      \mbox{
          \epsfxsize=7.50cm
           \epsffile{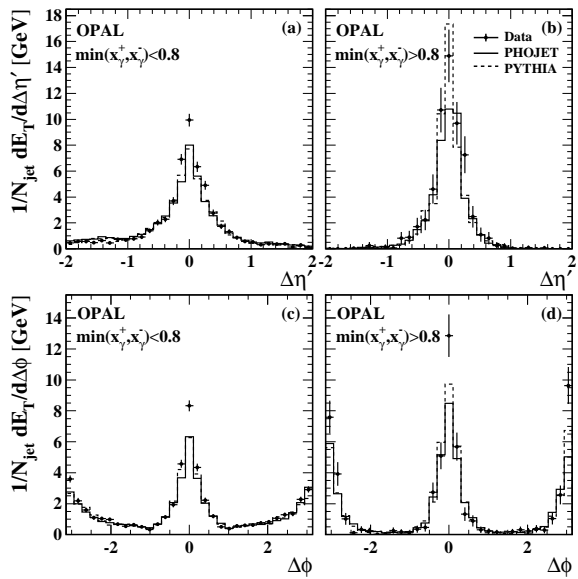}
           }
   \end{center}
\caption{Transverse energy flow 
measured relative to the direction of the jet in two-jet events. 
Jets from events with $\min(\xgp,\xgm)<0.8$  (a,c) 
and $\min(\xgp,\xgm)>0.8$  (b,d) are shown separately.
The energy flow is integrated over $|\Delta\phi|<\pi/2$
for the $\Delta\eta'$ projection (a,b) and
over $|\Delta\eta'|<1$ for the $\Delta\phi$ projection (c,d).
Statistical errors only are shown.}
\label{fig-xgflow}
\end{figure}

Figure~\ref{fig-xg} shows the number of two-jet events
as a function of $\xgp$, together with the
PYTHIA and PHOJET samples, after the detector simulation.
The $\xgm$ distribution, which is not shown, is consistent
with the $\xgp$ distribution within the statistical errors.
The peak expected for direct events at $\xgp=1$ 
is smeared out due to higher order QCD,
hadronisation and detector resolution effects.
Only statistical errors are
shown, since systematic effects largely
cancel in the ratio (Eq.~\ref{eq-xgpm}). 

The direct events, which are shown separately
for the PYTHIA sample, mainly contribute in the
region $\xgp>0.8$. 
Experimentally, samples with large and small direct contributions are 
separated by requiring 
$\min(\xgp,\xgm)>0.8$ and $\min(\xgp,\xgm)<0.8$, respectively. 
About 95~\% of all events in the
region $\min(\xgp,\xgm)>0.8$ originate from direct interactions
in PYTHIA.

Figure~\ref{fig-xgflow} shows the transverse energy flows
\begin{equation}
 \frac{1}{N_{\rm jet}}\frac{\mathrm{d}E_T}{{\mathrm d}(\Delta\eta')} \;\;\
\;\mbox{and}\;\;\;
 \frac{1}{N_{\rm jet}}\frac{\mathrm{d}E_T}{{\mathrm d}(\Delta\phi)}
\label{eq-flow}
\end{equation}
with respect to the jet direction for two-jet events with 
$\min(\xgp,\xgm)<0.8$ (low $\xg$)
and with $\min(\xgp,\xgm)>0.8$ (high $\xg$), separately.
The shape of the selected jet is seen clearly in both 
$\Delta\eta'$ and $\Delta\phi$.
No detector correction has been applied. The same tracks and clusters
which are used for jet finding are included, with
\begin{equation}\Delta\eta'=k(\eta-\etajet) \;\; \mbox{and}
\;\;\Delta\phi=\phi-\phijet.\end{equation} 
The rapidity difference is multiplied by a 
factor $k=\pm 1$.
The factor $k$ is chosen event-by-event to be
$k=+1$ for events with $\xgp>\xgm$ and $k=-1$ for
events with $\xgp<\xgm$.
Due to the choice of $k$, there is always more energy flow at $\Delta\eta'<0$.

In Fig.~\ref{fig-xgflow}, a
pedestal is observed which could indicate the existence of
a photon remnant. 
As expected, this enhancement in the region around
$|\Delta\phi|\approx\pi/2$ and at $|\Delta\eta'|>1$ is
more pronounced for low $\xg$ events.
The jets in high $\xg$ events  are much
more back-to-back in $\Delta\phi$ (Fig.~\ref{fig-xgflow}d)
than in low $\xg$ events (Fig.~\ref{fig-xgflow}c).
The pedestal in the $\Delta\phi$ region between the two jets at
$|\Delta\phi| \approx \pi/2$ is not observed in the high $\xg$ events.
Jets in high $\xg$ events are also observed, on average,
to have more average transverse energy and to be more collimated.
This is as expected for direct events, where all the available energy 
is used in the hard subsystem.
Both Monte Carlo models describe the transverse energy flow 
reasonably well, except for an
underestimate in the central region around the jet axis.  Nevertheless
the overall modelling is sufficiently good to justify using the Monte
Carlo models for unfolding the detector resolution effects in the 
jet cross-section measurements.

\section{Inclusive jet cross-sections}
\label{sec-cross}
\begin{figure}[htbp]
   \begin{center}
\hskip -0.2cm
      \mbox{
          \epsfxsize=7.55cm
           \epsffile{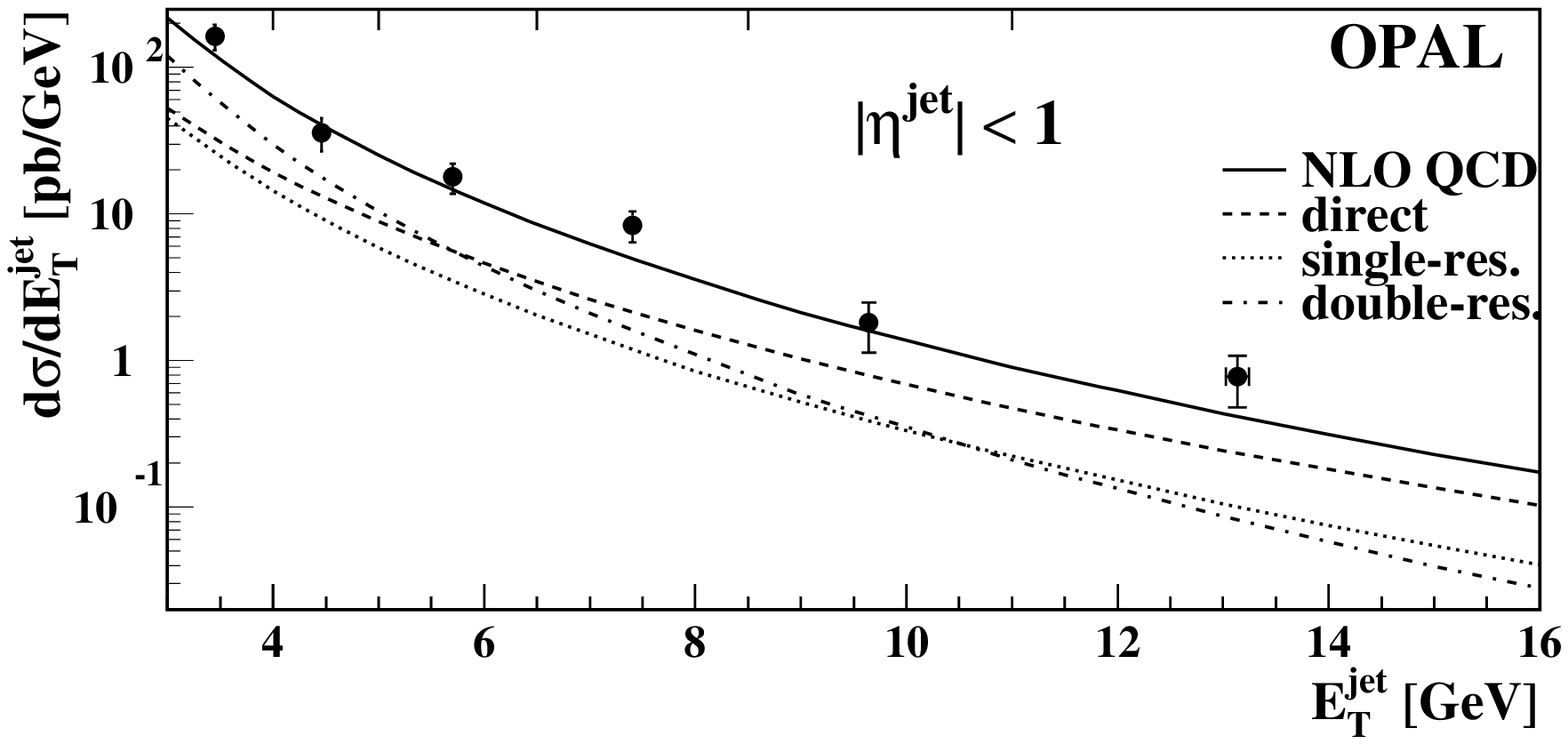}
           }
\hskip -0.2cm
      \mbox{
\hskip -0.2cm
          \epsfxsize=7.55cm
           \epsffile{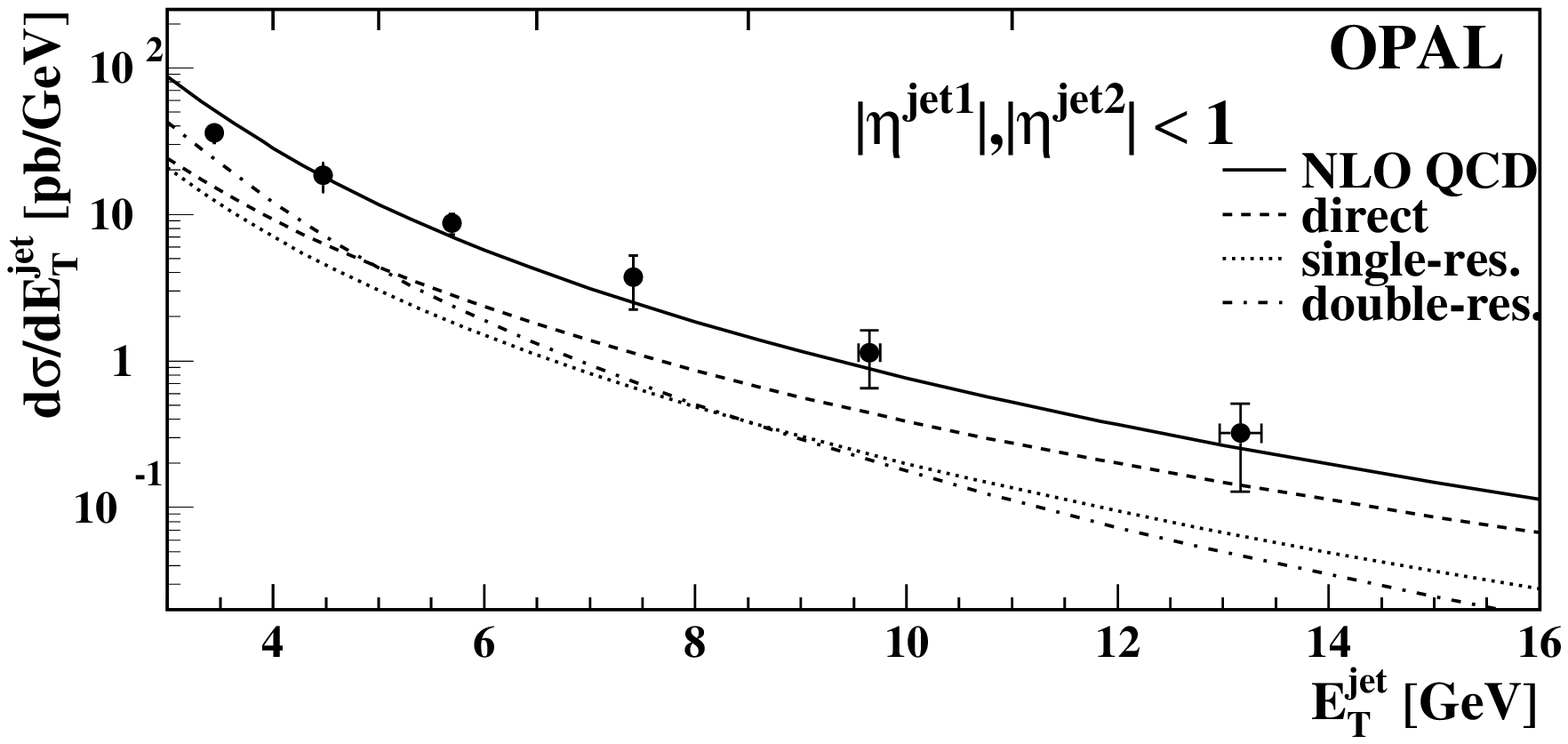}
           }
   \end{center}
\caption{The inclusive one-jet (upper plot) and two-jet (lower plot)
cross-sections as a function
of $\ETJET$ 
compared to the NLO calculation by 
Kleinwort and Kramer~\protect\cite{bib-kleinwort}.
The direct, single-resolved and double-resolved cross-sections
and the sum are shown separately. 
}
\label{fig-ptjet}
\end{figure}

The one- and two-jet cross-sections shown in 
Fig.~\ref{fig-ptjet} are corrected 
to the hadron level using the method of regularised unfolding.
The main systematic uncertainty on the jet cross-sections originates from
varying the energy scale of the ECAL 
in the Monte Carlo simulation by $\pm 5$~\% and 
from the unfolding procedure.
The model dependence of the unfolding is taken into
account by adding to the systematic error
the difference between the results obtained with 
PYTHIA, which are taken to be the central
values, and PHOJET.

The $\ETJET$ distribution is compared to a NLO
perturbative QCD calculation of the inclusive one-jet cross-section
by Kleinwort and Kramer~\cite{bib-kleinwort} who use
the NLO GRV parametrisation of the photon structure function~\cite{bib-grv}.
Their calculation was repeated for the kinematic conditions
of this analysis. All scales are chosen
to be equal to $\ETJET$. The strong coupling $\alpha_{\rm s}$ is
calculated from the two-loop formula with 
$\Lambda^{(5)}_{\overline{\rm MS}}=130$ MeV.
The ratio of the NLO to the LO inclusive one-jet cross-section
decreases from about 1.19 to 1.03 between $\ETJET=3$~GeV and
$\ETJET=16$~GeV.

The direct, single- and double-resolved parts of the one-jet cross-section
and their sum are shown separately. The
agreement between data and the calculation is good. The resolved
cross-sections dominate in the region $\ETJET\;\sleq\;5$~GeV,
whereas, at high $\ETJET$ the direct cross-section is largest.
The $\ETJET$ distribution
falls less steeply than expected in $\gamma$p and $\ppbar$ interactions,
because the fraction of hard interactions rises
from $\ppbar$ to $\gamma$p to $\gg$ interactions,
due to the additional direct photon interactions.
It should be noted that the NLO QCD calculation gives the jet cross-section
for massless partons, whereas
the jet cross-sections are measured for hadrons. 

The inclusive two-jet cross-section is measured
for events with at least $2$ jets using the two jets
with the highest $\ETJET$ in an event.
The $\ETJET$ distribution is also compared to the
NLO calculations~\cite{bib-kleinwort},
which have been extended to include the full NLO calculation for
the double-resolved two-jet cross-section.
The increase of the NLO compared to the LO inclusive
two-jet cross-section is less than 10~\% in the calculation.

\begin{figure}[htbp]
   \begin{center}
\hskip -0.25cm
      \mbox{
          \epsfxsize=7.58cm
           \epsffile{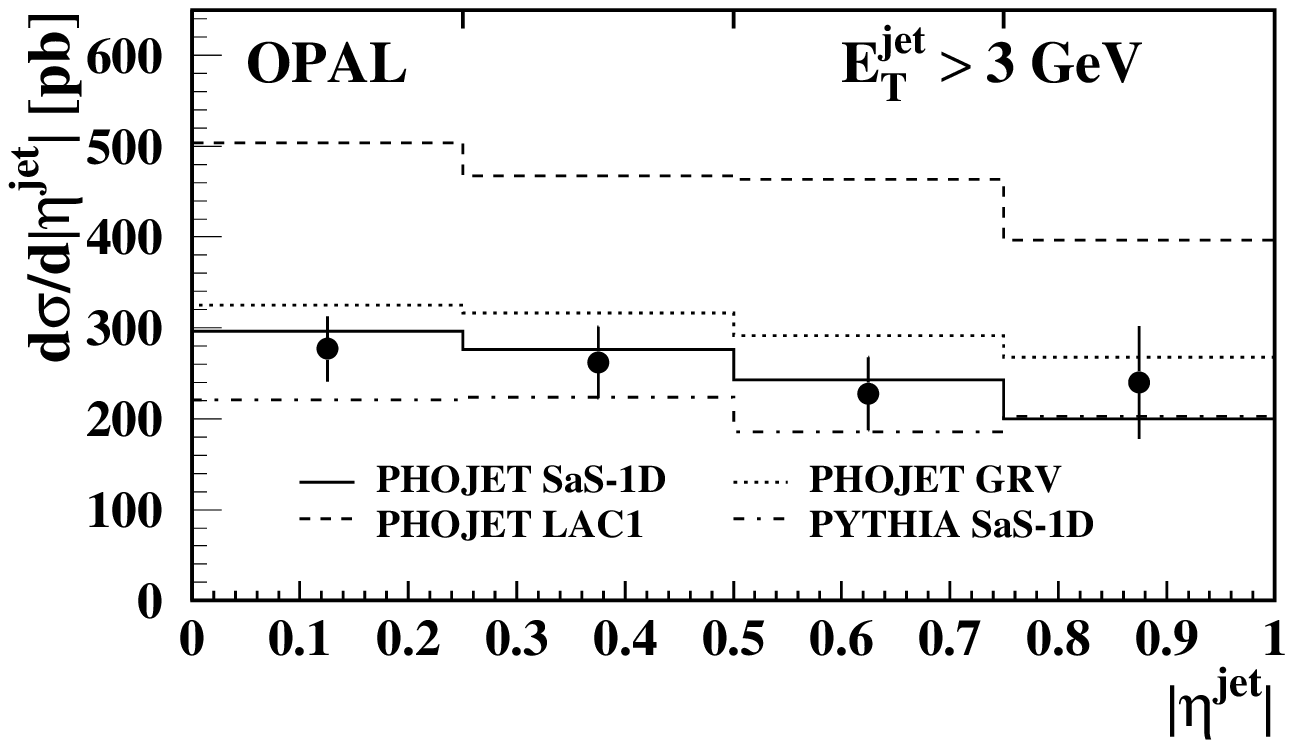}
           }
\hskip -0.25cm
      \mbox{
\hskip -0.25cm
          \epsfxsize=7.58cm
           \epsffile{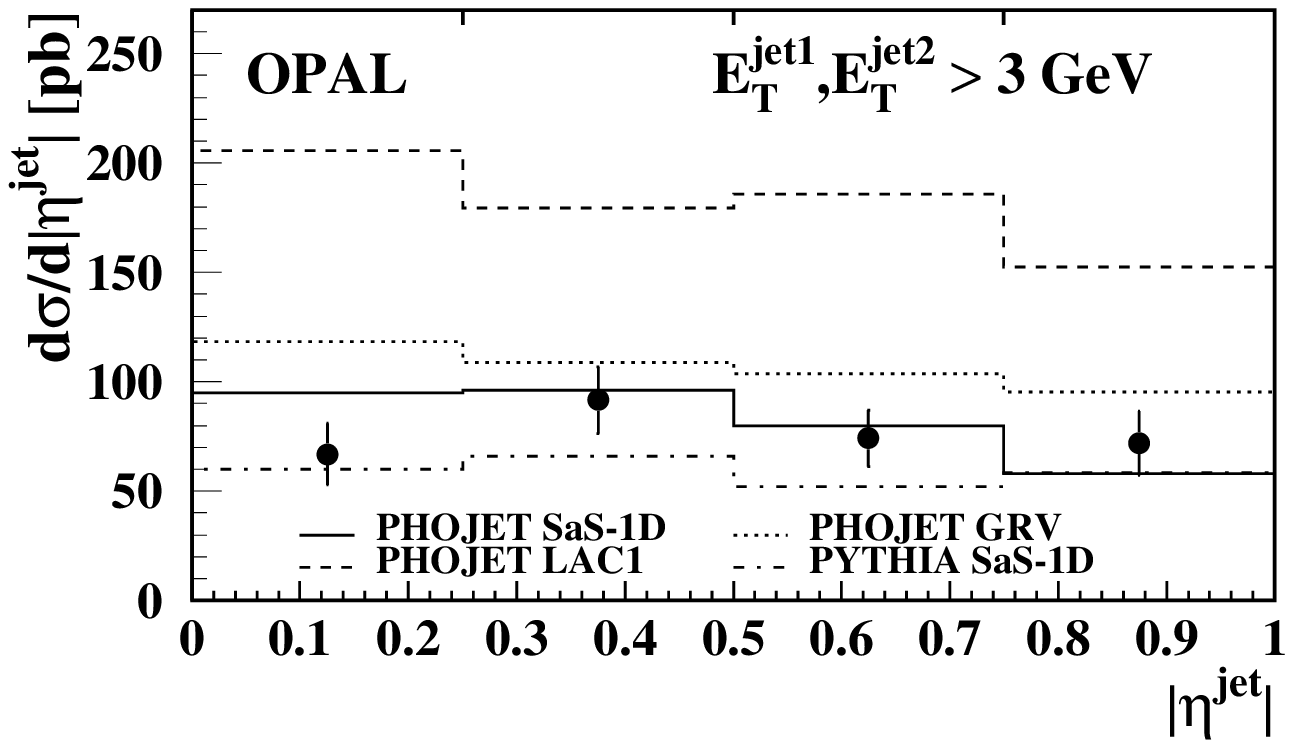}
           }
   \end{center}
\caption{The inclusive one-jet (upper plot)
and two-jet (lower plot) cross-sections as a function
of $|\etajet|$ compared to the LO QCD calculations of PYTHIA and PHOJET. 
}
\label{fig-etajet}
\end{figure}

The inclusive one- and two-jet cross-sections as a function of $|\etajet|$
are shown in Fig.~\ref{fig-etajet}.
Within the uncertainties
of the measurement, the distributions are nearly independent of $|\etajet|$ 
in agreement with the expectations of the Monte Carlo models. 

The total cross-sections, which are dominated
by the low $\ETJET$ events, depend on the photon structure function. 
In Fig.~\ref{fig-etajet} 
the total jet cross-sections predicted by the two Monte Carlo models
differ significantly even if the same photon structure function
(here SaS-1D) is used. 
This model dependence reduces the
sensitivity to the parametrisation of the photon structure function.
Different parametrisations were used as input to the PHOJET simulation. 
The GRV-LO and SaS-1D parametrisations describe the data
equally well, but the LAC1 parametrisation~\cite{bib-LAC1}
overestimates the total jet cross-section
by about a factor of two.

\section{Conclusions}
\label{sec-conclusions}
We have measured jet production in $\gg$ interactions with
the OPAL detector at $\sqee$ of 130 and 136 GeV, using a
cone algorithm with $R=1$,
$\ETJET>3$~GeV and $|\etajet|<1$.

Two-jet events originating from direct and resolved photon interactions
were separated experimentally using $\xgpm$. 
Jets in events with $\min(\xgp,\xgm)>0.8$  are expected to be produced mainly 
from direct interactions. These jets are observed to have, on average,
more average transverse energy and to be more collimated 
than jets in resolved events with $\min(\xgp,\xgm)<0.8$. In resolved events
a pedestal is observed in the transverse energy flows which
may be related to the photon remnant.
The Monte Carlo models PYTHIA and PHOJET describe 
the transverse energy flow distributions reasonably well.
 
The inclusive one-jet and two-jet production cross-sections
were measured as a function of $\ETJET$ and $|\etajet|$. 
The measurement extends the $\ETJET$ range of previous 
measurements~\cite{bib-amy} up to $\ETJET=16$~GeV. 
The $\ETJET$ dependent one- and two-jet cross-sections are in good agreement
with NLO QCD calculations~\cite{bib-kleinwort}.

Within the uncertainties of the measurements,
the jet cross-sections are nearly independent of $|\etajet|$ 
for $|\etajet|<1$. 
The total jet cross-section is dominated by the 
resolved cross-section.
Given the model dependence of the jet cross-sections in PYTHIA and PHOJET,
the GRV-LO and SaS-1D parametrisations describe the data equally well.
The LAC1 parametrisation overestimates the total jet cross-section
by about a factor of two.


\end{document}